\UseRawInputEncoding
\documentclass[aps,prl,twocolumn,floatfix,superscriptaddress]{revtex4-2} %linenumbers,

\usepackage{amsmath}  % needed for \tfrac, \bmatrix, etc.
\usepackage{amsfonts} % needed for bold Greek, Fraktur, and blackboard bold
\usepackage{graphicx} % needed for figures
\usepackage{epstopdf}
\usepackage{gensymb}
\usepackage{color}
\usepackage{soul}
\usepackage{comment}

\definecolor{DarkGreen}{RGB}{0,150,50}

\graphicspath{ {./MainFigures/} }
\usepackage[latin1]{inputenc}

\begin{document}
    %\title{ Intermittent Spontaneous Excitation Melts Acoustically Levitated Granular Rafts}
    \title{Hydrodynamic Coupling Melts Acoustically Levitated Crystalline Rafts}
    \begin{abstract}
% Immersed in fluid, particles interacting hydrodynamically often exhibit exotic, out-of-equilibrium behaviors that defy intuition. However, most of hydrodynamic interactions are associated with a prescribed symmetry breaking, either rotational or translational.Here we present a novel, acoustically levitated driven granular system that conserves symmetry on the global scale. We acoustically levitated a membrane of  particles, 27-45 microns in diameter. This quasi-2D membrane is a dissipative but driven granular system; the energy density of the acoustic cavity plays the role of an effective temperature and determines the magnitude of fluctuations in the system. At the length scales and frequencies considered here hydrodynamic flow between particles establishes interactions that are repulsive at a close range but attractive several diameters apart. Underdamped, hydrodynamically driven dynamics in this many-body system give rise to two kinetically distinct states: a crystal like and a liquid like state. We demonstrate the unique spatial kinetic energy structure in the liquid state arising from multi-body interactions. Furthermore, we show that the transition between the crystal and liquid state exhibits intermittent, spontaneous excitations of large-scale particle re-arrangement events.

The acoustic levitation of small particles provides a versatile platform to investigate the collective dynamical properties of self-assembled many-body systems in the presence of hydrodynamic coupling.
However, acoustic scattering forces can only generate attractive interactions at close range in the levitation plane, limiting self-assembly to rafts where particles come into direct, dissipative, contact.
Here, we overcome this limitation by using particles small enough that the viscosity of air establishes a repulsive streaming flow at close range. By tuning the size of particles relative to the characteristic length scale of the viscous flow, we control the interplay between attractive and repulsive forces.
In this novel granular raft, particles form an open lattice with tunable spacing.
Hydrodynamic coupling between particles gives rise to spontaneous excitations in the lattice, in turn driving intermittent particle rearrangements. Under the action of these fluctuations, the raft transitions from a predominantly quiescent, crystalline structure, to a two-dimensional liquid-like state. We show that this transition is characterized by dynamic heterogeneity and intermittency, as well as cooperative particle movements, that produce an effectively `cageless' crystal.
These findings shed light on fluid-coupling driven excitations that are difficult to isolate and control in many other hydrodynamic systems.
\end{abstract}
    
    \author{Brady Wu}%
    \email[email: ]{bwu34@uchicago.edu}
    \affiliation{Department of Physics, University of Chicago}
    \affiliation{James Franck Institute, University of Chicago}
    
    \author{Bryan VanSaders}%
    \affiliation{James Franck Institute, University of Chicago}

    \author{Melody X. Lim}%
    \thanks{Present address: Laboratory of Solid State and Atomic Physics, Cornell University}
    \affiliation{Department of Physics, University of Chicago}
    \affiliation{James Franck Institute, University of Chicago}
    % \affiliation{Laboratory of Solid State and Atomic Physics, Cornell University}
    % \affiliation{Kavli Institute at Cornell for Nanoscale Science, Cornell University}
    
    \author{Heinrich M.  Jaeger}
    \affiliation{Department of Physics, University of Chicago}
    \affiliation{James Franck Institute, University of Chicago}

    % \date{\today}

    \maketitle
    
    \let\clearpage\relax

Acoustic levitation has been widely used as a contact-free technique to manipulate particles and direct particle self-assembly \cite{marzo_holographic_2015,memoli_metamaterial_2017,foresti_acoustophoretic_2013,llewellyn-jones_3d_2016,melde_holograms_2016,ahmed_rotational_2016,shi_general_2019,chen_liquid_2017,tsujino_ultrasonic_2016}.
It has also been used to study the collective properties of many body systems \cite{lim_cluster_2019,lim_mechanical_2022,dong_colloidal_2019,abdelaziz_ultrasonic_2021}.
Within the nodal plane of an acoustic cavity, scattered sound waves establish tunable attractions between particles \cite{silva_acoustic_2014}, aggregating them together into a close-packed monolayer raft.
Sub-millimeter granular particles are too large for thermal noise to drive their dynamics. However, other athermal fluctuations can be introduced by slightly detuning the cavity excitation away from its resonant frequency, which  results in vertical oscillations \cite{rudnick_oscillational_1990, andrade_experimental_2019, lim_cluster_2019} or spontaneous rotation of the entire raft \cite{lim_mechanical_2022}.

Such fluctuations are a manifestation of hydrodynamic instabilities (HIs), where the configuration of particles moving relative to a surrounding fluid evolves unpredictably due to hydrodynamic coupling. 
HI-induced excitations in many-body systems can exhibit complex correlations that control collective properties, in contrast to uncorrelated thermal fluctuations.
In the above-mentioned prior work, the fluctuations emerged from feedback between a moving object and the acoustic cavity mode, an instability found already for a single levitated particle \cite{rudnick_oscillational_1990, andrade_experimental_2019}.  
A more subtle instability can arise from the hydrodynamic coupling among multiple particles.
For example, such HIs can occur among neighboring particles settling in a fluid \cite{janosi_chaotic_1997,chajwa_kepler_2019}, and it has been associated with the chaotic trajectories of small numbers of weakly interacting steel spheres in an oscillating glycerol bath \cite{thomas_structures_2004}.
Acoustic levitation has proven to be a powerful technique to observe multiple forms of HIs in systems of hundreds of particles evolving over long time scales.
However, strongly dissipative frictional particle contacts in prior studies \cite{lim_cluster_2019,lim_mechanical_2022} masked the presence of HIs arising from the relative motion of particles.

% \st{[b.w. the following sentence seems repetitive]Tunable control of the strength of particle contacts is needed, which is not possible with rafts where strong attractive forces always drive particles into direct contact }.
%\textcolor{blue}{\emph{HMJ:  I think the end of this very important paragraph needs to be a bit more explicit about the issues we are addressing and in motivating what follows. So I propose to replace the text starting with 'Despite....' by:}  Acoustic levitation offers an opportunity to observe this type of HI in systems containing more than a few interacting particles and to investigate the resulting cooperative effects in long-lived steady-states. However, the observation of such steady-states requires that the HIs are not masked by strongly dissipative frictional interactions, which was unavoidable in prior work on levitated rafts because attractive forces pulled particles into direct contact.}

\begin{figure*}
    \centering
    \includegraphics[width=2\columnwidth]{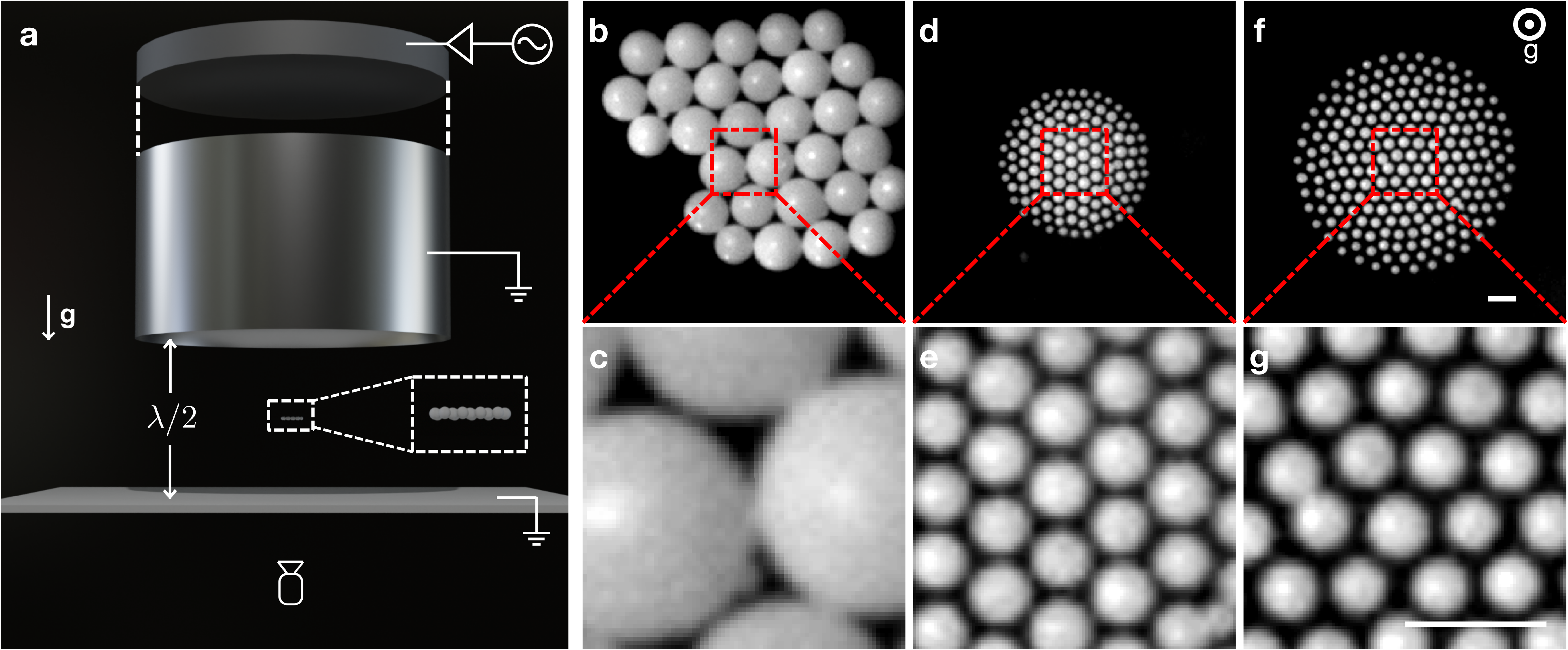}
    \caption{\textbf{Acoustically levitated particles with tunable interactions.}
    \textbf{a}, Schematic of the experimental setup.
    The transducer consists of an aluminum resonating horn (bottom) and a piezoelectric driving element (top).
    Ultrasonic standing waves (of wavelength $\lambda$) form between the transducer and reflector.
    Particles are levitated at the pressure nodal plane.
    A high-speed video camera is used to obtain the bottom view.
    \textbf{b}-\textbf{g}, Images of levitated particle clusters (top) and zoomed-in (bottom) views of interparticle spacing.
    Average particle diameter and frequency are \textbf{b} $D=180 \mu m, f=34.7$ kHz ($\lambda=9.55$ mm), \textbf{d} $D= 40 \mu m, f=64.3$ kHz ($\lambda=5.23$ mm), and \textbf{f} $D= 40 \mu m ,f=34.7$ kHz ($\lambda=9.55$ mm). Scale bars correspond to 100 $\mu m$.}
    \label{fig:fig1}
 \end{figure*}

Here, we overcome this limitation by levitating particles small enough that a viscosity-driven repulsive streaming flow can have an appreciable effect on the structure of the particle assemblies.
When the levitated particles are comparable in size to the extent of the streaming flows, the repulsive hydrodynamic forces can counteract attractive scattering forces at close range.
By tuning the particle size, and by extension, the relative strengths of attraction and repulsion, we are able to vary the stable distance between pairs of levitated particles from contact to a particle diameter apart (Fig \ref{fig:fig1} \textbf{b}-\textbf{f}). 
Underdamped and contactless, the particles can be now be driven by excitations arising from their hydrodynamic coupling.

%Due to its non-pairwise nature, systems driven by HIs can display many exotic dynamics unexpected from fluid-free systems.
We find that the sound energy density of the cavity determines the magnitude of HI-driven excitations, controlling the transition from a quiescent crystal-like state to a highly diffusive liquid-like state.
As a result of fluid-induced spatio-temporal correlations, this transition is mediated by intermittent, avalanche-like dynamics where a single displacement event can trigger system wide rearrangements, a clear distinction from the spatially homogeneous thermal melting transition \cite{zahn_dynamic_2000,thorneywork_two-dimensional_2017}.
Fundamentally, the avalanche-like excitation arises as a consequence of the correlated motion between neighboring particles starting from the ballistic time scale, which produces an effectively `cageless' crystal over long time scales.
In contrast to similar phenomenon in kinetically trapped systems \cite{hwang_understanding_2016}, avalanche-like events in levitated rafts do not relax by the descent of a rough configurational landscape and can continue indefinitely.
% \note{This is an interesting comment - explain more, and connect to the intermittency?}
% While long-ranged interactions between air flows and the levitated raft give rise to a finite-size effect, the above results exhibit the same trends for only the bulk particles. 
% It is important to note that this hydrodynamically -derived perturbation does not arise from any predefined in-plane driving such as per-particle rotation or self propulsion \cite{Bililign2022, Wagner2017}. 
 
Our experimental setup is illustrated in Fig \ref{fig:fig1} \textbf{a}.
A resonating acoustic cavity is driven on one side by an ultrasonic transducer, which is bolted to an aluminum horn to maximize power output.
The resonant frequency $f_0$ of the coupled transducer and horn is 34,870Hz, with an associated free space wavelength $\lambda = c/f_0 \simeq 9.8mm$, where $c$ is the speed of sound in air (Supplementary Section 1).
Here, we use polyethylene particles of diameter $D = 30-60 \mu m$ (Supplementary Section 2 and 3), such that the setup operates deep in the Rayleigh scattering regime ($\lambda > D$).
Unlike larger levitated objects \cite{rudnick_oscillational_1990, andrade_experimental_2019, lim_cluster_2019}, these particles compose a negligible volume fraction of the resonating cavity. As a result, changing the driving frequency or amplitude adjusts the energy density of the acoustic cavity without triggering spontaneous vertical oscillations of individual particles. 
Confined by single-scattering forces (i.e. primary forces) \cite{king_acoustic_1934, gorkov_forces_1962}, particles levitate at the pressure nodal plane, and are imaged from below using a high-speed camera (Fig \ref{fig:fig1} \textbf{b}-\textbf{f}).

\begin{figure}
    \centering
    \includegraphics[width=\columnwidth]{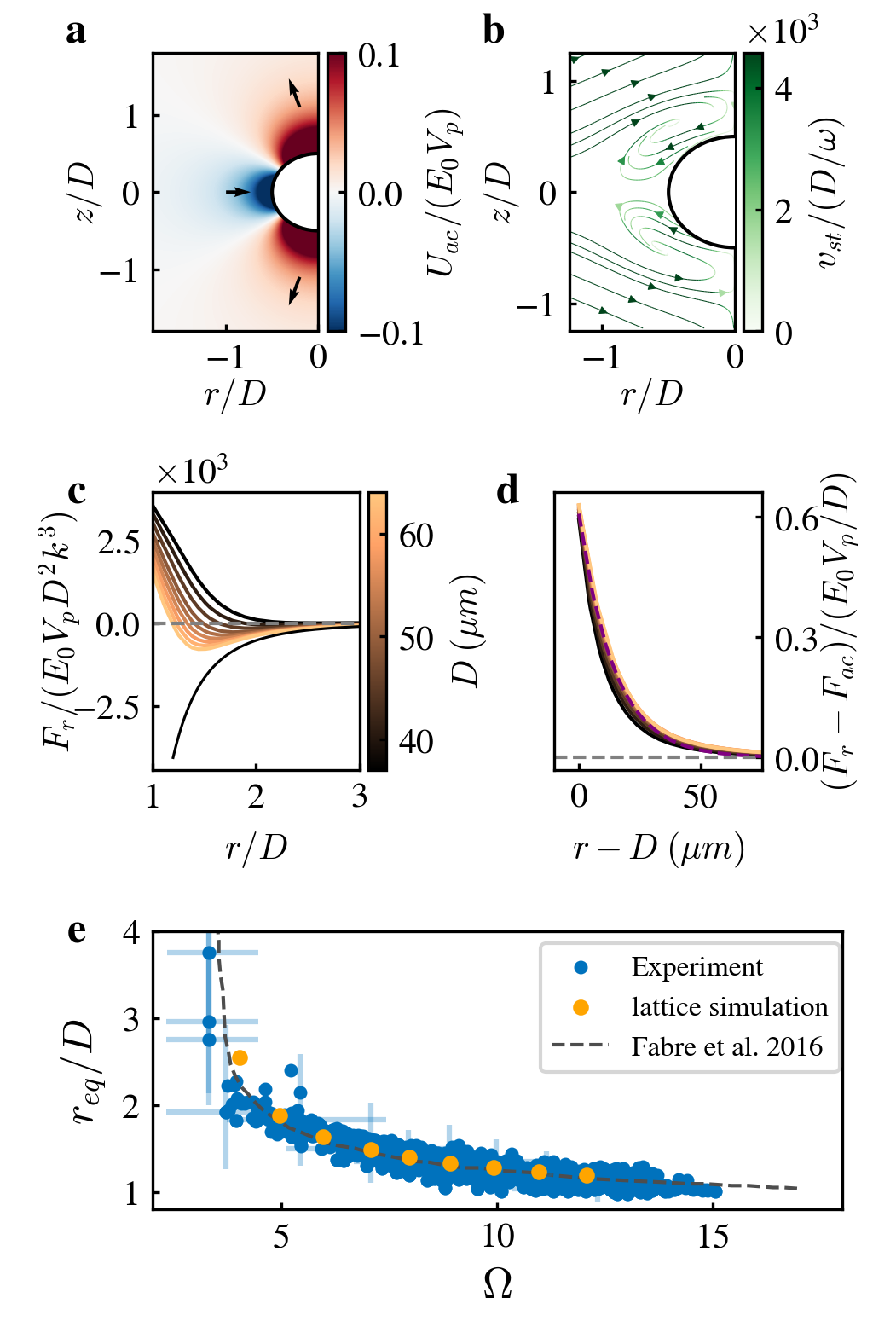}
    \caption{\textbf{Short-ranged repulsive interactions from viscous effects.}
    \textbf{a}, The acoustic scattering potential, $U_{ac}$, in the x-z plane.
    The interaction is attractive in the equatorial while repulsive in the polar direction.
    The white circle represents the particle.
    \textbf{b}, Time-averaged microstreaming flow of magnitude $v_{st}$ around a sphere, as obtained by Lane \cite{Lane1955}.
    The direction of flow-induced forces in the levitation plane are opposite to that in \textbf{a}.
    \textbf{c}, The magnitude of force in the equatorial direction $F_r$ as a function of radial distance $r$ for various particle diameter, as calculated with a lattice fluid simulation.
    The black curve represents the force from the scattering expansion following \cite{silva_acoustic_2014}.
    \textbf{d}, The magnitude of the microstreaming repulsive interaction, approximated by subtracting the scattering expansion from the lattice fluid force traces in \textbf{c}. The purple dashed line represents an exponential fit, $a*exp(-(r-D)/\kappa)$; $\kappa \approx 14 \mu m$.
    \textbf{e}, The equilibrium distance $r_{eq}/D$ between a particle pair as a function of Stokes number $\Omega$.
    Blue errorbars represent the uncertainties for selected data points (Supplementary Section 2).}
    \label{fig:fig2}
\end{figure}

\emph{Tunable attractive and repulsive interactions.}
At low energy density in the cavity, particles levitate in stationary configurations and do not fluctuate.
The equilibrium distances between particles are determined by the competition of attractive and repulsive interparticle forces.
Attractive forces arise from the scattering of sound between particles.
These long-range secondary radiation forces are anisotropic, with attraction for particle configurations in the nodal plane and repulsion out of the plane \cite{silva_acoustic_2014} (Fig.~\ref{fig:fig2} \textbf{a}).
The magnitude of in-plane attraction as a function of distance scales as 
 
 \begin{equation}
 	F_{rad}(r) \propto -\frac{ E_0 D^6 }{r^4 \lambda^{3}},
	\label{eq:Frad}
 \end{equation}
 where $E_0 \equiv \rho_0 V_0^2/2$ is the energy density of the acoustic field in air with maximum velocity $V_0$, density $\rho_0$, wavelength $\lambda$, and $r$ is the center-to-center distance between particles of diameter $D$ \cite{silva_acoustic_2014}.
 
 Interparticle repulsion arises from a viscosity ($\nu$) effect, and is not captured by the above scattering analysis.
 Near particle surfaces a viscous boundary layer of characteristic thickness $\delta_\nu=\sqrt{\nu/\pi f}$ forms, over which the velocity of air relative to the particle surface decays to zero.
 Within the boundary layer, acoustic energy is coupled to steady, short-ranged flows.
 For particle sizes much larger than $\delta_\nu$, the influence of this boundary layer is negligible; however, for $D \approx \delta_\nu$, the time-averaged, non-linear microstreaming flows that develop within the viscous boundary layer surrounding the particle become significant.
 These flows form two toroidal vortices around the sphere, aligned with the axis of the exciting wave \cite{lane_acoustical_1955}.
 Flow in the levitation plane near the particle surface is outward, opposing attraction from the radiation force (Fig.~\ref{fig:fig2} \textbf{b}).
 The magnitude of this microstreaming-induced repulsion depends primarily on $\delta_\nu$, which is similar for particles of all sizes at the same driving frequency $f$.
 Therefore, as particle size decreases, the relative magnitude of the viscous correction to the short-ranged interaction grows.
 As viscous forces are nearly constant, while $F_{rad}$ scales with $D^6$ (Eq.~\ref{eq:Frad}), interparticle separation increases with smaller particle size.

To analyze this quantitatively, we calculate interparticle forces using the Lattice Boltzmann method (LBM), a discretized direct fluid simulation technique (see Methods).
Figure \ref{fig:fig2} \textbf{c} shows the force experienced between pairs of variously-sized particles, scaled by the non-dimensional prefactor of the scattering expansion pair interaction.
For $r/D>2$, pair forces are attractive for most sizes, and larger particles more closely follow the form of the scattering expansion (black curve).
We estimate the form of the viscous repulsive interaction by subtracting the scattering expansion from the net force, yielding the data in Fig. ~\ref{fig:fig2} \textbf{d}.
As expected, the decay of viscous repulsion from the particle surface is nearly independent of diameter, and exponential over the thickness of the viscous layer $\delta_\nu$.
%, in accordance with our hypothesis. 

The competition between scattering and viscous forces produces stable in-plane equilibrium separations which are controlled by the Stokes number $\Omega=2\pi f R^2/\nu$ \cite{fabre_acoustic_2017}.
Figure~\ref{fig:fig2} \textbf{e} compares the numerical results for the equilibrium, in-plane center-to-center particle distance $r_{eq}$ from Fabre et al.~\cite{fabre_acoustic_2017} with our LBM simulations and experimental observations, showing close agreement.
%At separations larger than 1.5$D$, the force minima become very shallow, which makes observation challenging. 
%Nevertheless, within imaging error the experiments track the numerics well. 
Importantly, we find no significant difference in $r_{eq}$ when comparing isolated particle pairs with pairs of neighboring particles inside large rafts (Supplementary Section 4). 
%Our experimental observations of isolated particle pairs as well as neighboring particles in rafts (both in blue) are consistent with LBM simulations and prior numerical results by Fabre et al. \cite{Fabre2017}, as shown in (Fig.~\ref{fig2} \textbf{e}).
These results are consistent with prior investigations of centimeter-scale spherical particles in a confined, oscillating viscous liquid~\cite{klotsa_chain_2009,voth_ordered_2002,thomas_structures_2004,otto_measurements_2008}.
%Importantly, such studies were restricted to systems in close contact with a confining wall, a limitation that acoustic levitation transcends.

  \begin{figure}
    \centering
    \includegraphics[width=\columnwidth]{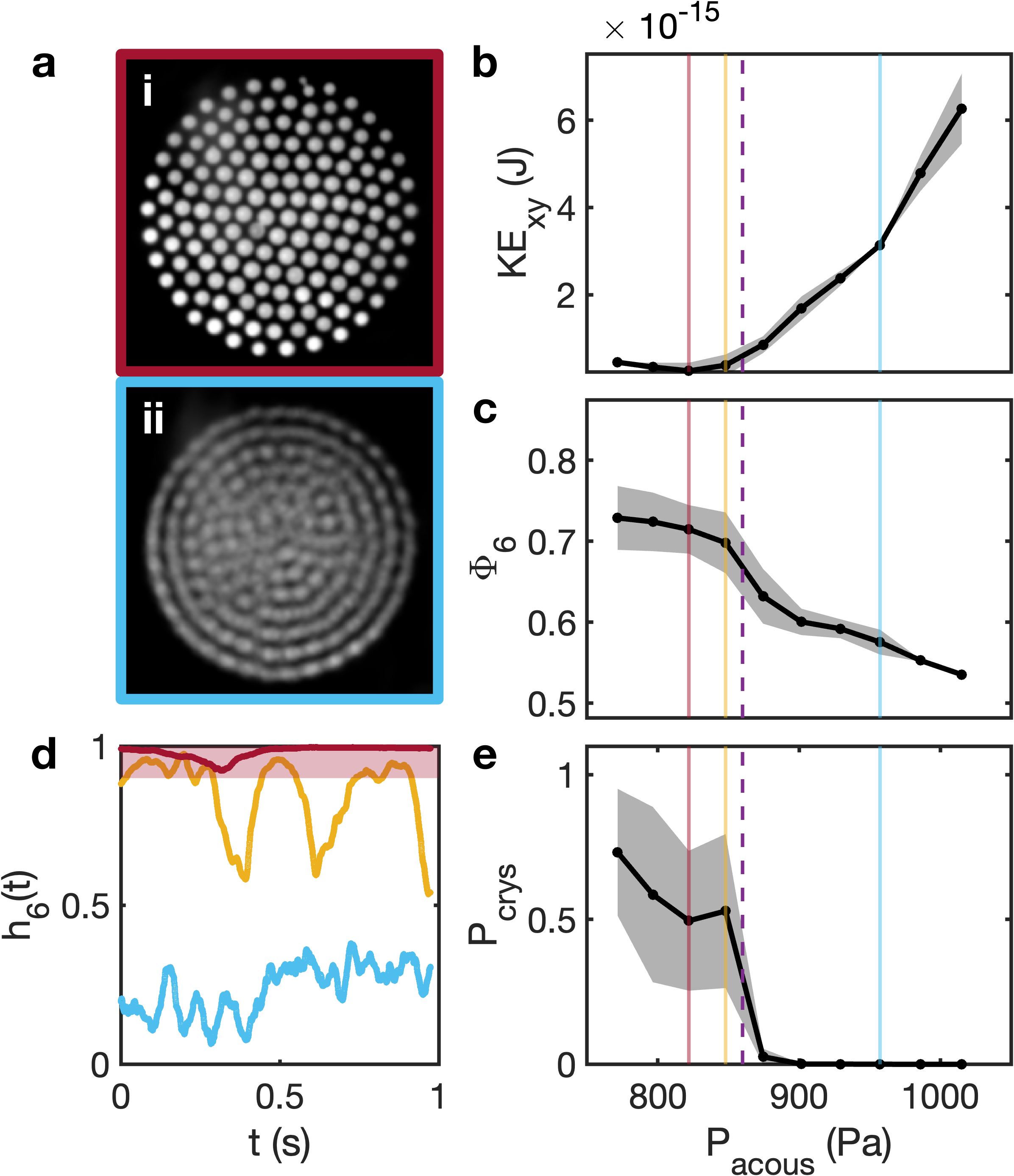}
    \caption{\textbf{Acoustic energy density controls crystal-liquid transition.}
    \textbf{a}, Time averaged images of crystal-like (\textbf{i}) and liquid-like (\textbf{ii}) states over $167ms$ at $P_{acous}=822\,Pa$ and $P_{acous}=957\,Pa$.
     \textbf{b}-\textbf{c}, The average kinetic energy in xy direction $KE_{xy}$, and the bond order parameter $\phi_6$ as a function of acoustic pressure, $P_{acous}$. 
     \textbf{d}, The bond correlation function $h_6(t)=\langle \phi_6(t-\Delta t) \cdot \phi_6(t)\rangle $ for crystal (maroon), liquid (blue), and intermittent dynamics (orange). $\Delta t=10ms$.
     We define $P_{crys}$ as the probability of $h_6>0.9$ (the shaded region).
     \textbf{e}, The probability of crystal-like state, $P_{crys}$, as a function of $P_{acous}$.}
    \label{fig:three_states}
 \end{figure}

\emph{Order-to-disorder transition.}
% \note{This section may flow better if these two sentences are deleted, and go straight into levitated rafts -- the stochastic dynamics are very interesting, and deserve to be introduced with your results. Experiments~\cite{Thomas2004} \edits{on} particles in an oscillating liquid also observed instability-driven excitations capable of driving stochastic dynamics. 
% While these experiments were limited to just a few interacting particles, acoustically levitated rafts provide an opportunity to explore how interparticle HIs affect configurations comprising much larger particle numbers. }
%Can such mechanisms energize acoustically levitated rafts in the streaming regime?
In the following we focus on levitated rafts with $\Omega \approx6$, and induce stochastic particle dynamics by increasing the applied oscillation amplitude, parameterized here by the acoustic pressure in the cavity, $P_{acous}$.
The intensity of acoustic driving controls the transition from a quiescent crystal-like state (Fig.~\ref{fig:three_states} \textbf{a}, \textbf{i}) with well-defined equilibrium interparticle spacings, to a liquid-like state (Fig. \ref{fig:three_states} \textbf{a}, \textbf{ii}) with rapidly fluctuating particle positions and associated kinetic energy (Fig. \ref{fig:three_states} \textbf{b} and Supplementary Video 1). 
We note that all motion in this liquid state occurs within the levitation plane and does not involve excursions into the vertical direction (Supplementary Video 2).
The particles' velocity distribution is isotropic and well-described as a traditional granular gas (Supplementary Section 5).
The interaction of the raft with the acoustic field in the cavity produces a reduced local energy density at the edge of the raft, thus producing distinctive rings of particles that rarely leave their respective annuli (Supplementary Section 6).

To quantify how disorder emerges in the crystal-like state as $P_{acous}$ is increased, we calculate the spatially and long-time averaged bond order parameter $\Phi_6= \langle \frac{1}{N_{nn}}\sum_{nn} e^{i6\theta_i}\rangle_N $, where $\theta_i$ is the angle between the bond and an arbitrary axis, $nn$ denotes nearest neighbor, $N_{nn}$ is the number of neighbor particles, and the average is over all particles $N$ \cite{zahn_dynamic_2000,thorneywork_two-dimensional_2017}. 
We find that $\Phi_6$ monotonically decreases with increasing $P_{acous}$ (Fig. \ref{fig:three_states} \textbf{c}). 
At first glance this is surprising, given that the strength of interparticle, in-plane attractive forces increases with energy density in the cavity and thus $P_{acous}$ (see Eq. 1 and Fig. ~\ref{fig:fig2} \textbf{c}).
Stronger interparticle binding promotes crystalline order; however, we find that the acoustic energy density increases the strength of HI-driven perturbations more rapidly, allowing us to to observe an order-to-disorder transition.

% Our results suggest that increasing~$P_{acous}$ has a stronger effect on the inter-particle HI driven perturbations than on the in-plane attractive forces (which would tend to increase crystalline order). 
% As a consequence, we are able to observe an order-to-disorder transition, driven by HI perturbations. \note{This is a neat observation! Is there some scaling ansatz we can get for the HI with $P_{acous}$? Maybe something even as simple as ``HIs are second order in the velocity field so we expect quadratic scaling with energy density"?}
%driven by cavity energy density, instabilities must scale as a function of $P_{acous}$ more rapidly than stabilizing, conservative interparticle forces. 

\begin{figure*}
    \centering
    \includegraphics[width=2\columnwidth]{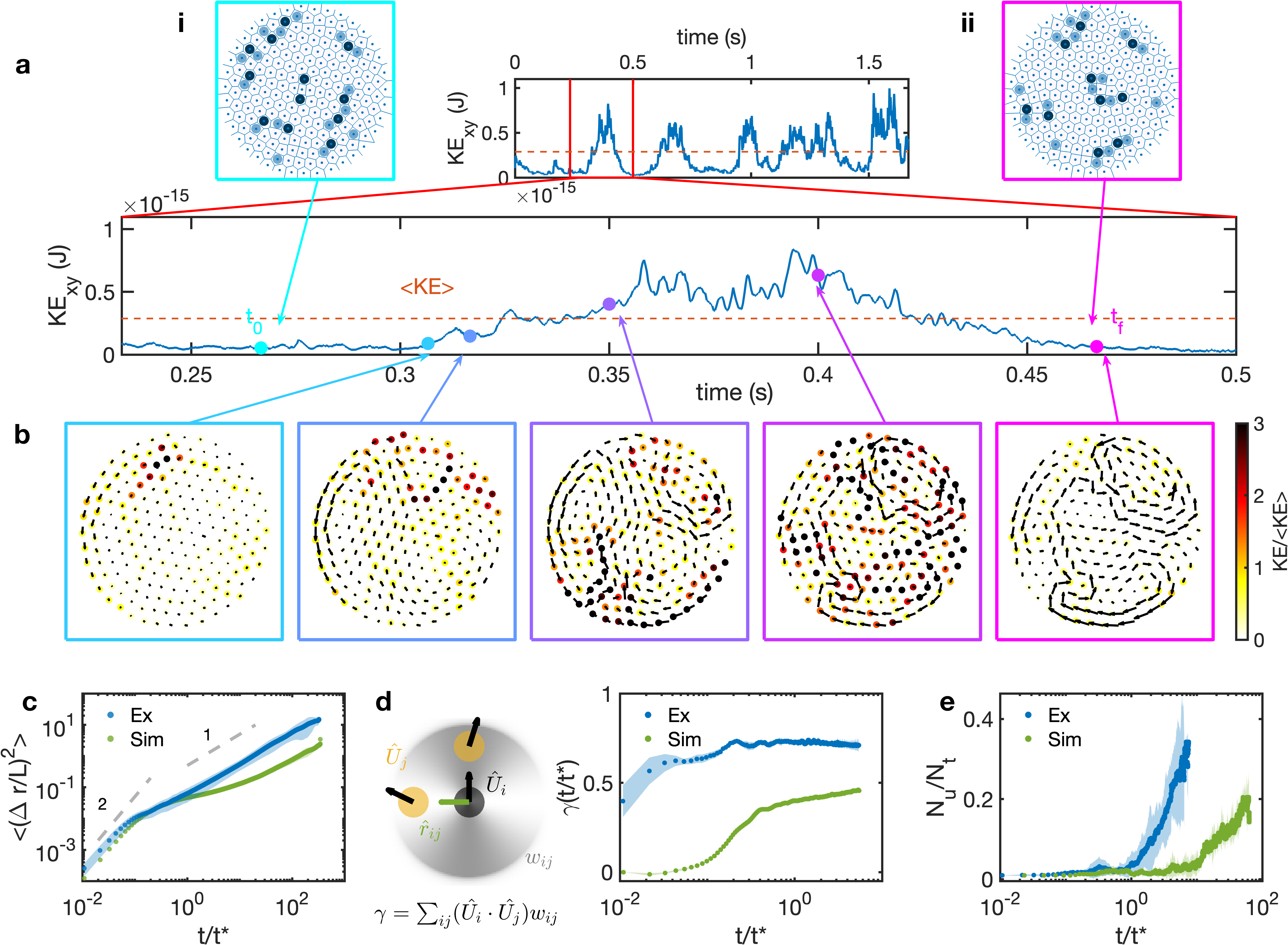}
    \caption{\textbf{Cooperative uncaging drives intermittent dynamics.}
    \textbf{a}, The kinetic energy per particle within a raft during an avalanche.
    The inset exhibits several avalanches within a span of 1.5s.
    The particular avalanche in the main panel of \textbf{a} is boxed in red.
    The average kinetic energy per particle over 10s is plotted in orange dashed lines.
    The marked points correspond to the snapshots in \textbf{b}.
    \textbf{i}-\textbf{ii}, The voronoi construction of the cluster configuration at $t_0$ and $t_f$. Light and dark blue represents the 5 and 7 particles defect. 
    \textbf{b}, A time sequence of a raft displaying an intermittent avalanche event.
    Arrows denote the displacement field from time $t_0$ before the event, as indicated in \textbf{a}.
    Particles are colored with their current values of $KE/\langle KE\rangle $.
    \textbf{c}, Normalized mean square displacement $\langle (\Delta r/L)^2\rangle $ as a function of $t/t^*$ for both the experiment and MD simulation.
    L denotes the lattice spacing, and $t^*=L/\langle v\rangle $.
    \textbf{d}, $\langle \gamma\rangle $ as a function of the normalized time.
    The schematic shows the definition of $\gamma$.
    $w_{ij}=(\hat{U}_i \cdot \hat{r}_{ij})^2$ is the angular weighting function and depicted with the shaded area; the directions parallel to the particle motion are weighted more heavily.
    \textbf{e}, The fraction of uncaged particles in a raft, $N_u/N_t$, as a function of$t/t^*$.
    $N_u$ and $N_t$ are the uncaged and total particle number, respectively. 
    A particle is considered uncaged if the displacement from time $t_0$ is greater than $\langle v\rangle  t^*$. The shaded area represents standard deviation from 10 independent avalanche events in the same raft.}
    \label{fig:avalanche}
\end{figure*}

We observe that this order-to-disorder transition is characterized by sudden, avalanche-like rearrangements (Fig.~\ref{fig:avalanche} \textbf{a}).
To quantify these excursions, we calculate a spatially averaged, local order correlation function, $h_6(t,\Delta t)=\langle \Phi_6(t-\Delta t) \cdot \Phi_6(t)^*\rangle_N$, which measures how much local 6-fold orientational order ($\Phi_6$) changed over a short time interval $\Delta t$ before time $t$, averaged over $N$ particles.
$h_6=1$ indicates no change in ordering, while $h_6$ close to zero corresponds to the decorrelation of local bond order.
A characteristic decorrelation time $\Delta t =10 ms$ was chosen for the liquid state, but the behavior of $h_6$ shown here is not sensitive to choices of $\Delta t$ from 5 to 100ms.
% \note{this could make for a useful supplementary figure?}

Figure \ref{fig:three_states} \textbf{d} shows the evolution of $h_6$ for three representative cases.
 Rafts exhibit intermittent dynamics (orange), where $h_6(t)$ switches between low and high values at random intervals, indicating that the raft behaves as a crystal-like state with stochastic bursts of liquid-like rearrangements. 
 The quiescent, crystal-like state has a wide range of lifetime, from fractions of a second to a few seconds, but will eventually be interrupted by a liquid-like excitation. 
In the crystal-like state the particle displacements are smaller than a lattice spacing, and account for all decorrelations $h_6>0.9$ (red shaded band).
For example, the dip in $h_6$ for the red trace around 0.3s is due to phonon motion in an otherwise stationary crystal.
% While this is similar to the intermittent excitation observed in dusty plasma system, the excitation here is driven by interparticle HIs rather than vertical oscillations \cite{gogia2017}. \note{This seems like a bit of a tangent - consider moving to the conclusion?}
Finally, at large $P_{acous}$, the raft enters a liquid-like state, with $h_6<0.5$.
Defining the probability $P_{crys}$ of finding the raft in the crystal-like state as the fraction of time that $h_6>0.9$, we observe a sharp drop around $P_{acous}^*\approx 860 Pa=1.1 P_{acous}^{min}$ ($P_{acous}^*$ is the purple dashed line in Fig.~\ref{fig:three_states} \textbf{b},\textbf{c},\textbf{e}, $P_{acous}^{min}=780$Pa is the minimum pressure at which rafts can be stably levitated), signaling the transition from the stochastically interrupted crystal-like state, to a consistently liquid-like state. 
%This behavior starkly contrasts with equilibrium order-to-disorder transitions, in which the ordered state approaches melting through temporally homogeneous fluctuations.

% The system melts into a liquid-like state for $P_{acous}>P^*$ (Fig. \ref{fig:three_states} \textbf{e}), where $P^*\approx 860 Pa$ (purple dashed line). While the interior of the liquid states exhibits many chain-like rearrangements, the edge does not seem to participate; the outer two-layer of particles only slowly move in the azimuthal direction, forming the ring-like structure in Fig. \ref{fig:three_states} \textbf{b} inset. It is also important to note that parallel to the direction of motion, the neighboring particles seem to move together, thus forming a chain of comoving particles.  

%The observation that fluid energy density determines the magnitude of fluctuation is common across many diverse HCSs \cite{Bililign2022,Thomas2004}, where the fluctuations are driven primarily through hydrodynamic instabilities. However, while the classical thermal fluctuations are decorrelated both in space and time, there is no reason to expect the same for hydrodyanmically driven fluctuation. 

%\vspace{5mm} \noindent \textbf{Intermittent dynamics arising from cooperative motion}

% The relation between HIs and local acoustic energy density sets the conditions that allow for the stochastic, avalanche-like events responsible for intermittent switching between quiescent crystal-like and liquid-like states (Fig~\ref{fig:three_states}\textbf{d},orange).
% Fluid-body interaction not only introduces global acoustic energy structure but may also locally drives the HIs between particles, which is most apparent when rafts display intermittent dynamics, switching between quiescent crystal-like and liquid-like states.
\emph{Intermittent dynamics from cooperative motion.} We now look in more detail at the structural rearrangements that are produced by HIs inside a raft.
%The order-to-disorder transition, mediated by intermittent excitation, could potentially reveal the nature of HIs induced fluctuation. 
Intermittent excursions into the liquid-like state can be identified by pulses in the kinetic energy per particle (Fig. \ref{fig:avalanche} \textbf{a} inset), where each pulse is an intense, avalanche-like excitation.
Between such events, the system settles into an idle, highly ordered crystal state (Supplementary Video 3). 
These excited and quiescent states exist for similar amounts of time and typically each last for a fraction of a second. 
At the onset of an event (Fig. \ref{fig:avalanche} \textbf{a}), the kinetic energy increases smoothly at about the same rate as it decays by viscous damping of individual particle motions in air.
This similarity in the rates of energy increase and decrease implies that the energy is injected locally via viscous streaming flows, on a per particle basis, rather than through a sudden external, boundary-coupled perturbation (e.g., via an air current).
% Despite this, external perturbations are still expected to play a key role in initiating small configurational changes that grow into avalanche-like events.

To better understand these intermittent excursions, we next examine the per-particle dynamics of a representative event.
We define a reference time $t_0$, immediately before the onset of the event for comparison with later times, $t$.
In Fig.~\ref{fig:avalanche} \textbf{b}, arrows indicate particle displacements from $t_0$ to $t$.
Particles positions at time $t$ are shown and colored according to their kinetic energy.
An event typically starts with chain-like rearrangements at the raft edge, the locations of which are correlated with defects that delineate the interface between the raft's core and its outer layers, as shown in Fig. \ref{fig:avalanche} \textbf{a}, \textbf{i}--\textbf{ii} (Supplementary section 7) \cite{kong_topological_2003}.
Particle motion creates more defects which propagate through the entire extent of the raft (Supplementary section 8), inducing an avalanche-like burst of collective rearrangement that persists until the kinetic energy is dissipated and the system attains some stable configuration.
We note that there is significant spatial heterogeneity in the particle rearrangements during an event, with the most mobile particles tending to collectively move in a linear path through the raft, similar to what has been observed in glassy supercooled liquids \cite{widmer-cooper_how_2004,keys_measurement_2007,reichhardt_fluctuating_2003}. 

A crucial distinction between our levitated rafts and supercooled liquids is in the pre- and post- event configurations.
Glassy systems drive avalanche events by descending a rough free-energy landscape, a process that starts with large displacements localized near `hot spots', but eventually arrests as the ground state is approached.
In contrast, levitated rafts undergo intermittent avalanche-like events indefinitely, and crystal-like configurations before and after events are not distinctly different (Fig. \ref{fig:avalanche} \textbf{a}, \textbf{i}--\textbf{ii} and Supplementary section 9).
Furthermore, avalanche-like events in levitated rafts are not localized and eventually involve the entire system. 
% \note{Earlier (end of last paragraph) it was argued that the events are very spatially heterogeneous, but now it is stated that the events include the entire system. Is there a fine-grained distinction that I'm missing? This comparison also seems a little out-of-place -- is this something that belongs at this point in the text? It could be abbreviated to a single sentence that reads ``In contrast to rearrangements in supercooled liquids, which undergo avalanche events by descending a rough free energy landscape, the levitated rafts rearrange indefinitely, without much change in the configuration of particles".}

In order to illuminate the nonequilibrium nature of these hydrodynamically driven raft rearrangements, we compare our experimental results with predictions based on a thermal ensemble of particles. 
To this end, we plot the mean square displacement (MSD) for both the experiment and a molecular dynamics simulation (Supplementary section 10) with pair-wise interactions (Fig. \ref{fig:avalanche} \textbf{c}).
To properly compare the data, we non-dimensionalize distances by the lattice spacing $L$, and times by the time taken for a particle to ballistically travel a distance of one lattice spacing $t^*=L/\langle v\rangle_{N,t}$.
For particles in a thermal bath and with $k_BT/V_B = 0.04$ (the case most analogous to our experiments, see Supplementary section 10 for details), where $V_B$ is the binding energy, the MSD traces exhibit ballistic behavior for $t<<t^*$, caging for $t\approx t^*$, and diffusive motion for $t>>t^*$ \cite{weeks_subdiffusion_2002}.
In contrast, the experimental MSD does not exhibit any caging behavior.
Instead, it transitions directly from ballistic to diffusive behavior around $t\approx 0.1 t^*$.
This behavior is reminiscent to that of a liquid or gas, without lattice ordering to induce intermediate-time caging. 
Notably, although the experiment has an effective temperature $\langle KE_{x,y} \rangle/V_B = 8\times 10^{-4}$, nearly 50 times lower than the simulation counterpart, the experimental particles are more mobile due to the presence of intermittent avalanches.

To further examine this exotic behavior, we define the velocity-direction correlation parameter
\begin{equation}
    \gamma(t)=\sum_{ij}(\hat{U_i}(t) \cdot \hat{U_j}(t)) w_{ij}(t) ,
\end{equation}
where $\hat{U_i}$ and $\hat{U_j}$ are the unit vectors of displacement for neighboring particles $i$ and $j$, and $\hat{r}_{ij}$ is the displacement vector between particles. In addition, we define a weight density $w_{ij}(t)=(\hat{U}_i(t) \cdot \hat{r}_{ij}(t))^2$, indicated by the shading in the schematic of Fig. ~\ref{fig:avalanche} \textbf{d}, such that directions parallel to the velocity of the particle are weighted more than those perpendicular. 
Our results are plotted in Fig. ~\ref{fig:avalanche} \textbf{d}, and show that the levitated particles display highly correlated motions even at short times. 
In contrast, the motions of particles in the simulation are uncorrelated, and only begin to correlate after the ballistic regime $t>0.1 t^*$.
Fundamentally, this contrast highlights the difference between excitation induced by HIs and by thermal fluctuation.
Excitations from HIs display a high degree of spatio-temporal correlation and depend on the configuration of nearby particles. 

We show that this spatio-temporal correlation results in the lack of particle caging by measuring the number of particles that escape their caging neighborhood over time. Here, particles are considered to have escaped their cage if they traversed at least one lattice spacing relative to $t_0$. Our results are plotted in Fig. \ref{fig:avalanche} \textbf{e}, and show that the levitated particles begin to uncage at $t\approx t^*$, which is the time particles will take to cross a distance $L$ if their motion is ballistic. 
In contrast, the simulated system only starts to uncage for $t\approx 10 t^*$, which is when the particle motion becomes diffusive. 
This distinction shows how the correlated motions make the levitated crystal susceptible to perturbations, as a single uncaging event is amplified into a system-spanning event by the tendency of interparticle HIs to correlate particle motions.

Taken together, our results establish acoustically-levitated rafts as a potent platform for investigating how hydrodynamic coupling can spontaneously excite a many-body system.
By tuning the relative strength of attractive scattering interactions and repulsive microstreaming flows, particle configurations exhibit hydrodynamic instabilities that scale in magnitude with applied acoustic pressure more quickly than any stabilizing acoustic binding.
As a consequence, increasing acoustic pressure drives rafts through an order-to-disorder transition.
Unlike the melting of a two-dimensional crystal in thermal equilibrium, this order-to-disorder transition occurs through intermittent periods of liquid-like behavior that eventually merge into a two-dimensional fluid-like steady state.
These intermittent, spatially correlated events reveal that hydrodynamic-instability-driven excitations are not equivalent to their thermal counterparts, in line with recent results for active systems that demonstrate the central role of statistical decorrelation between driven and fluctuating degrees of freedom in order to recover thermal-like behavior \cite{han_fluctuating_2021}.
Levitated rafts may also serve as a highly suitable system to explore experimentally the link between spatially-varying dissipation and dynamic transitions, as seen in simulations of active matter \cite{cagnetta_large_2017, tociu_how_2019}.
Finally, our results provide valuable insight into the limits of acoustic many-body control, elucidating a key instability mechanism that affects the precision of acoustic manipulation at small scales.

    \newpage
    \section{Experimental Setup and Method}
\subsection{Experiment}
Our acoustic cavity is driven on one side by a commercial ultrasonic transducer.
An aluminum horn is bolted onto the transducer to amplify the acoustic power generation.
The bottom of the metallic horn has a concave radius of curvature $R=50mm$ and was painted black to minimize light reflection.
The concave geometry weakly confines the particles at the center of the trap.
A commercial heat tape, regulated with a PID temperature controller, is wrapped around the horn and maintains the transducer at $35\pm 0.5^\circ$C for experimental repeatability.
The transducer is driven by applying a sinusoidal wave of peak-to-peak voltage $V_{pp}$ 
(60-200V) and frequency close to the resonance frequency of the horn, $f_0=34.85kHz$; this signal is generated from a function generator and amplified by a high-voltage amplifier (A-301 HV amplifier, AA Lab Systems).
The reflector consists of an acrylic plate (bottom) and an ITO-coated glass slide (top). To reduce the effect of tribocharging, both the horn and reflector were grounded.
The distance between the edge of the horn and the reflector can be adjusted with a translation stage to $\lambda/2=4.9mm$.
In order to minimize the effect of air currents, the entire setup was enclosed within an acrylic box with dimensions much larger than the levitation region ($l\times w \times h=61\times 30 \times 46 \times  cm^3$).

We used polyethylene spherical particles (Cospheric, material density $\rho=1000kgm^{-3}$, diameter d=30-60 $\mu m$, see Supplementary Section 2 and 3 for particle size distribution).
The particles are stored, and all experiments are performed within a humidity and temperature-controlled laboratory environment (45-50 \% relative humidity,22-24$ ^\circ$C). 

Before and after each experiment, the setup was cleaned with de-ionized water and dried with compressed air.
For each experiment, the particles were added to the cavity with a tweezer.
Videos are recorded with a high-speed camera (Vision Research Phantom T1340) at 3000 frames per second.
The acoustic pressure generated is measured with Xarion optical microphone (Eta100 Ultra).

\subsection{Lattice Boltzmann Simulation}

The Lattice Boltzmann method (LBM) was used to perform simulations of particles interacting through acoustically-mediated forces.
This method captures the full extent of the fluid-structure interactions, naturally including the effects of viscous dissipation and momentum transfer due to multiple scattering events.
LBM simulations of the acoustic cavity were performed within the \texttt{waLBerla} framework \cite{bauer2021walberla}.
A single relaxation time scheme with a viscosity matching that of air was used \cite{yu2003visc}.
Particles were simulated in periodic domains to avoid spurious boundary reflections.
A spatially-homogeneous background force field was used to oscillate the background fluid.
The \texttt{PE} functionality of the \texttt{waLBerla} framework was used to simulate the interaction of particles with the acoustic field~\cite{gotz2010pe}.
Hydrodynamic forces between the particles and fluid were handled with the partially-saturated cells method \cite{owen2011efficient}.
A local cell size of~$D/15$, where~$D$ is the particle diameter, was sufficiently high lattice resolution for accurate force calculations.
% excitation, bistable media, actively driven
% Intermittent spontaneous excitation melts an acoustically levitated granular raft
    \newpage
    \bibliography{library}
    \hfill \break
\noindent\textbf{Acknowledgments}\\
We thank Nina Brown, Tali Khain, Qinghao Mao, and Severine Atis for useful and inspiring discussions.
This research was supported by the National Science Foundation through award number DMR-2104733.
The work utilized the shared experimental facilities at the University of Chicago MRSEC, which is funded by the National Science Foundation under award number DMR-2011854. The research also benefited from computational resources and services supported by the Research Computing Center at the University of Chicago.

\hfill \break
\noindent\textbf{Author contributions}\\
B.W., B.V.S., and H.M.J.~designed the study.
M.X.L.~designed the original experimental apparatus.
B.W.~modified the apparatus and performed the experiments.
B.W.~performed molecular dynamics simulations and analyzed the data.
B.V.S.~performed the lattice Boltzmann simulations and analyzed the data.
All authors interpreted the results and and contributed to writing the manuscript.
%\st{B.W. designed and performed experiments and MD simulations and analyzed data. B.V. designed and performed LBM simulations. M.X.L. designed experiments. H.M.J. supervised research. All authors discussed the results and wrote the paper.}

\hfill \break
\noindent\textbf{Competing interests}\\
The authors declare no competing interests.
\end{document}